\documentclass[a4paper,11pt]{article}
\usepackage{pos}
\usepackage{makecell}

\title{Forward silicon vertex/tracking detector design and R$\&$D for the future Electron-Ion Collider}

\author*[a]{Xuan Li}
\author[a]{Melynda Brooks}
\author[a]{Matt Durham}
\author[a]{Yasser Corrales Morales}
\author[a]{Astrid Morreale}
\author[a]{Christopher Prokop}
\author[a]{Eric Renner}
\author[a]{Walter Sondheim}

\affiliation[a]{Los Alamos National Laboratory,\\
  Los Alamos, NM, USA}

\emailAdd{xuanli@lanl.gov}

\abstract{The proposed high-luminosity high-energy Electron-Ion Collider (EIC) will provide a clean environment to precisely study several fundamental questions in the fields of high-energy and nuclear physics . A low material budget and high granularity silicon vertex/tracking detector is critical to carry out a series of hadron and jet measurements at the future EIC especially for the heavy flavor product reconstruction or tagging. The conceptual design of a proposed forward silicon tracking detector with the pseudorapidity coverage from 1.2 to 3.5 has been developed in integration with different magnet options and the other EIC detector sub-systems. The tracking performance of this detector enables precise heavy flavor hadron and jet measurements in the hadron beam going direction. The detector R$\&$D for the proposed silicon technology candidates: Low Gain Avalanche Diode (LGAD) and radiation hard depleted Monolithic Active Pixel Sensor (MALTA), which can provide good spatial and timing resolutions, is underway. Bench test results of the LGAD and MALTA prototype sensors will be discussed.}

\FullConference{%
  *** Particles and Nuclei International Conference - PANIC2021 ***\\
  *** 5 - 10 September, 2021 ***\\
  *** Online ***
}

\begin{document}
\maketitle

\section{Introduction}
The proposed Electron-Ion Collider (EIC) aims to solve several fundamental questions in the field of high-energy nuclear physics. An example is exploring the hadronization process \cite{eic_YR}. The EIC  will utilize polarized and unpolarized electron+proton ($e+p$) and electron+nucleus ($e+A$) collisions with a variety of nuclear species (mass number A = 2-208) at the center of mass energy from 20 GeV to 141 GeV. The expected instantaneous luminosity is around $10^{33-34} cm^{-2}sec^{-1}$, which is a factor of 1000 higher than the Hadron-Electron Ring Accelerator (HERA) collider. The beam crossing rate varies from 1 $ns$ to 10 $ns$ depending on the collision system. A high granularity, low material budget tracking detector, which can provide good spatial and momentum resolutions, is required especially in the hadron beam going direction to precisely measure charged particles produced by the asymmetric collisions at the EIC. A Forward Silicon Tracker (FST) based on the Monolithic Active Pixel Sensor (MAPS) technology has been developed to provide precise track reconstruction in the pseudorapidity region of 1.2 to 3.5. A series of heavy flavor hadron and jet observables have been enabled by this detector and will provide enhanced sensitivities to explore the hadronization process in vacuum and in a nuclear medium \cite{lanl_eic, lanl_fst}. The conceptual design of the proposed FST, its tracking performance and progresses of the associated detector R$\&$D will be discussed.

\section{The Proposed Forward Silicon Tracker Conceptual Design and Its Performance}
\begin{figure}[!htb]
\begin{center}
\includegraphics[width=0.3\textwidth]{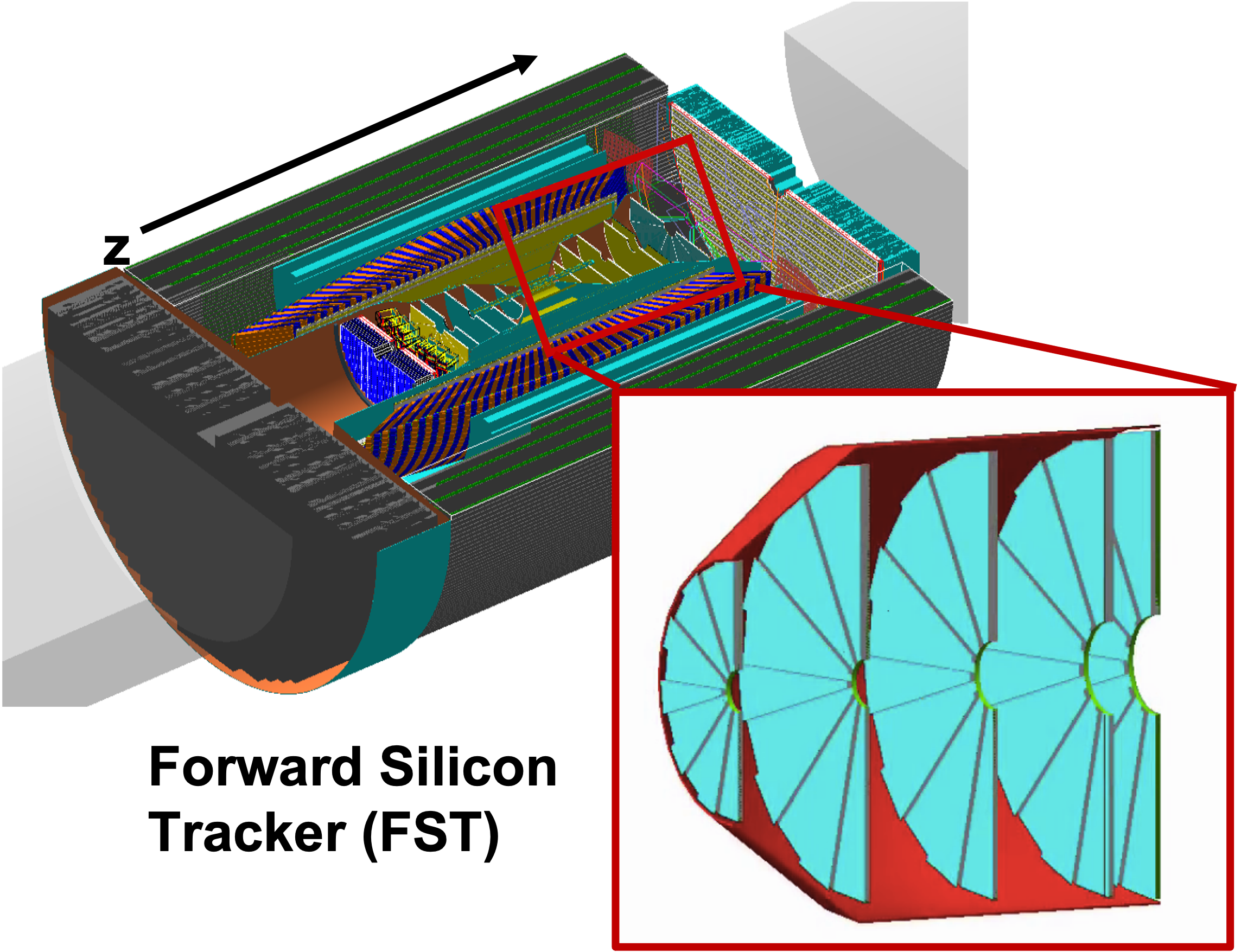}
\includegraphics[width=0.3\textwidth]{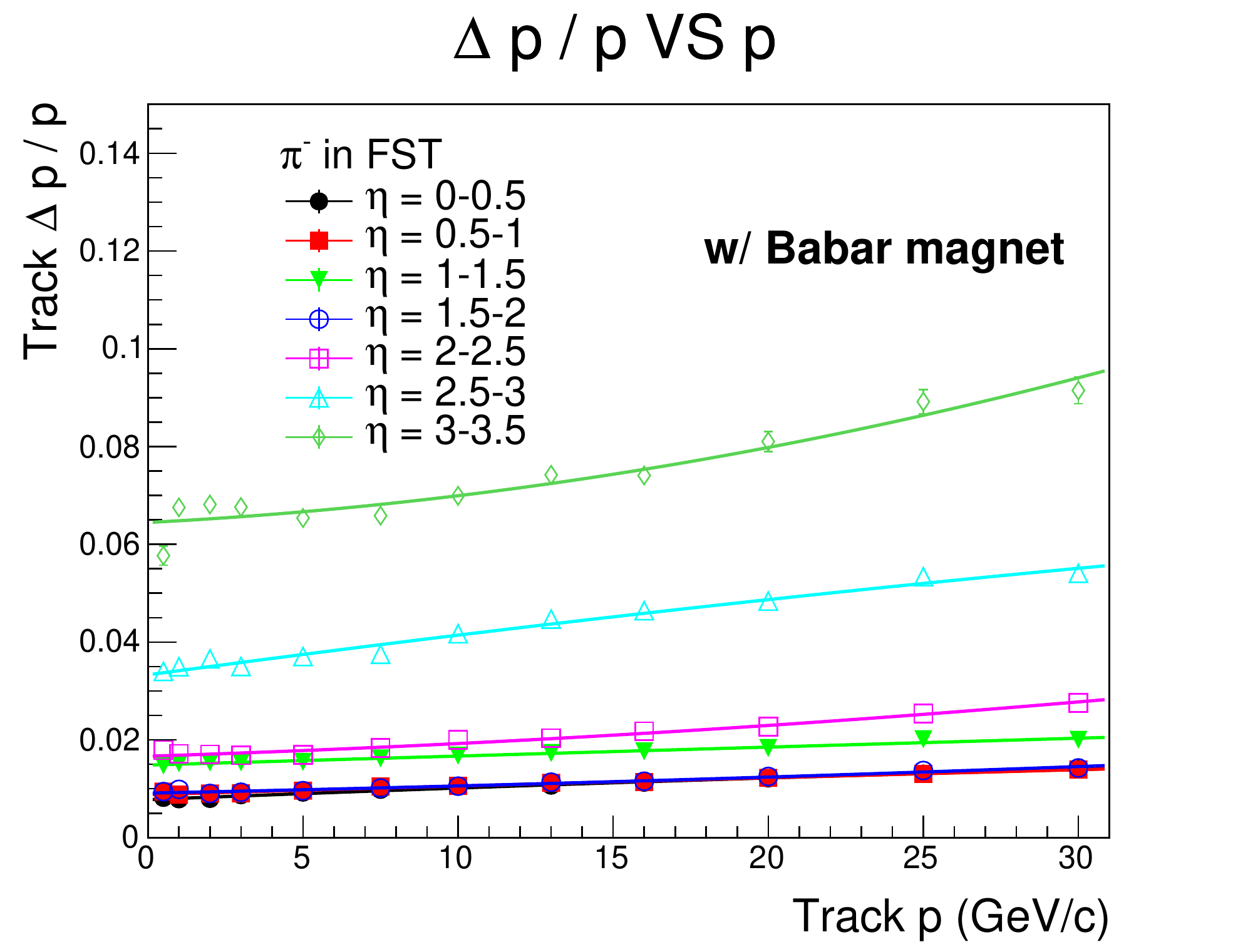}
\includegraphics[width=0.3\textwidth]{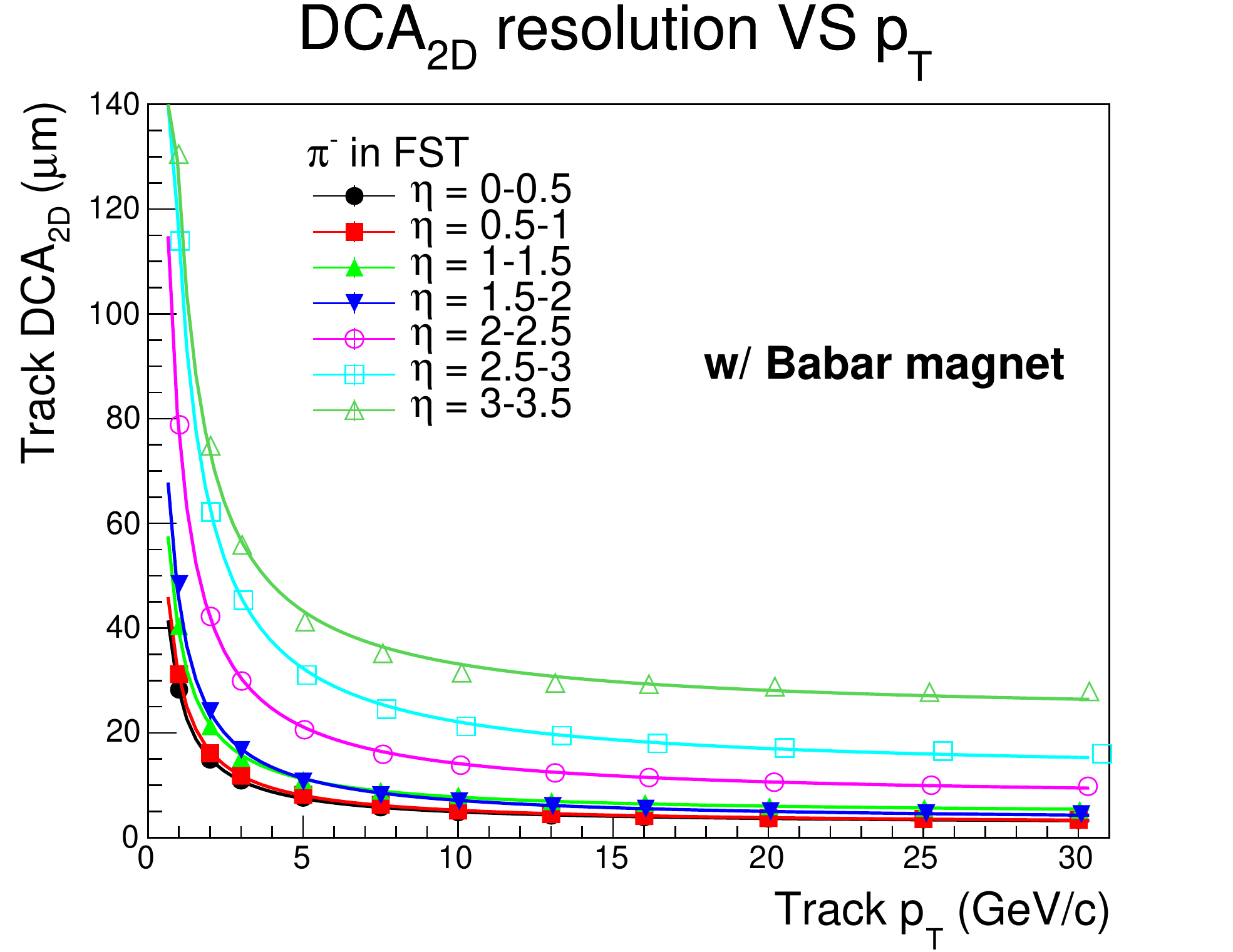}
\end{center}
\caption{The FST conceptual design implemented in GEANT4, which is integrated with the other EIC detector subsystems is illustrated in the left. The tracking momentum dependent momentum resolutions and the transverse momentum dependent transverse Distance of Closet Approach ($\text{DCA}_{\text{2D}}$) resolutions in different pseudorapidity regions of $0 \leq \eta \leq 3.5$ with the Babar magnet are shown in the middle and right.}
\label{fig:fst_per}
\end{figure}

The current design of the proposed FST consists of three MAPS \cite{maps} disks with the pixel pitch of 10 $\mu$m and two Depleted MAPS (referred to as MALTA) \cite{malta} disks with the pixel pitch of 36.4 $\mu$m. In addition to good momentum and spatial resolutions which can be offered by general MAPS trackers \cite{maps}, this hybrid silicon tracker provides fast timing resolution at around 5 $ns$, which can help separate events belonging to different beam bunch crossings at the EIC and suppress backgrounds. The latest geometry parameters of the FST conceptual design are listed in Table \ref{tab:fst_geo}. This detector concept has higher pixel densities compared to existing silicon detector at Relativistic Heavy Ion Collider (RHIC) and the Large Hadron Collider (LHC). The detailed detector layout, which includes the silicon sensor segmentation, cooling system, readout distribution and mechanical structure has been implemented in the GEANT4 simulation. 
\begin{table}[ht]
\small
\begin{center} 
\begin{tabular}{|c|c|c|c|c|c|c|c| } 
 \hline
 \makecell{ Disk \\ index }& technology & inner radius & outer radius & z location & pixel pitch & \makecell{ sensor \\ thickness } &  \makecell{ No. of \\ pixels } \\ \hline
1 & MAPS & 3.5 cm & 15.5 cm & 30 cm & 10 $\mu$m & 50 $\mu$m & $\approx$0.72B \\ \hline
2 & MAPS & 3.5 cm &  33.5 cm & 53.75 cm & 10 $\mu$m & 50 $\mu$m & $\approx$3.49B \\ \hline
3 & MAPS & 5.1 cm & 41.1 cm & 77.5 cm & 10 $\mu$m & 50 $\mu$m & $\approx$5.22B \\ \hline
4 & MALTA & 7.0 cm & 45.0 cm & 106.25 cm & 36.4 $\mu$m & 100 $\mu$m & $\approx$0.39B \\ \hline
5 & MALTA & 8.0 cm & 46.0 cm & 125 cm & 36.4 $\mu$m & 100 $\mu$m & $\approx$0.46B \\ \hline
\end{tabular}
 \caption{The FST conceptual design geometry \cite{lanl_fst}.}
    \label{tab:fst_geo}
\end{center}
\end{table}
The top panel of Figure~\ref{fig:fst_per} shows the geometry of the proposed FST integrated with the other EIC detector subsystems inside the Babar magnet, which provides the maximum magnetic field at 1.4 T. The track momentum dependent momentum resolutions and transverse momentum dependent transverse Distance of Closet Approach ($\text{DCA}_{\text{2D}}$) resolutions in seven different pseudorapidity ranges from 0 to 3.5 are illustrated in the right panel of Figure~\ref{fig:fst_per}. Different FST designs and the corresponding tracking performances from different integrated detector and magnet options have been studied \cite{ lanl_fst}. Recent simulation studies have validated that this proposed FST can enable precise heavy flavor hadron and jet measurements in the forward pseudorapidity region of $1.2<\eta<3.5$ at the EIC \cite{lanl_eic,lanl_fst}.

\begin{figure}[!htb]
\begin{center}
\includegraphics[width=0.86\textwidth]{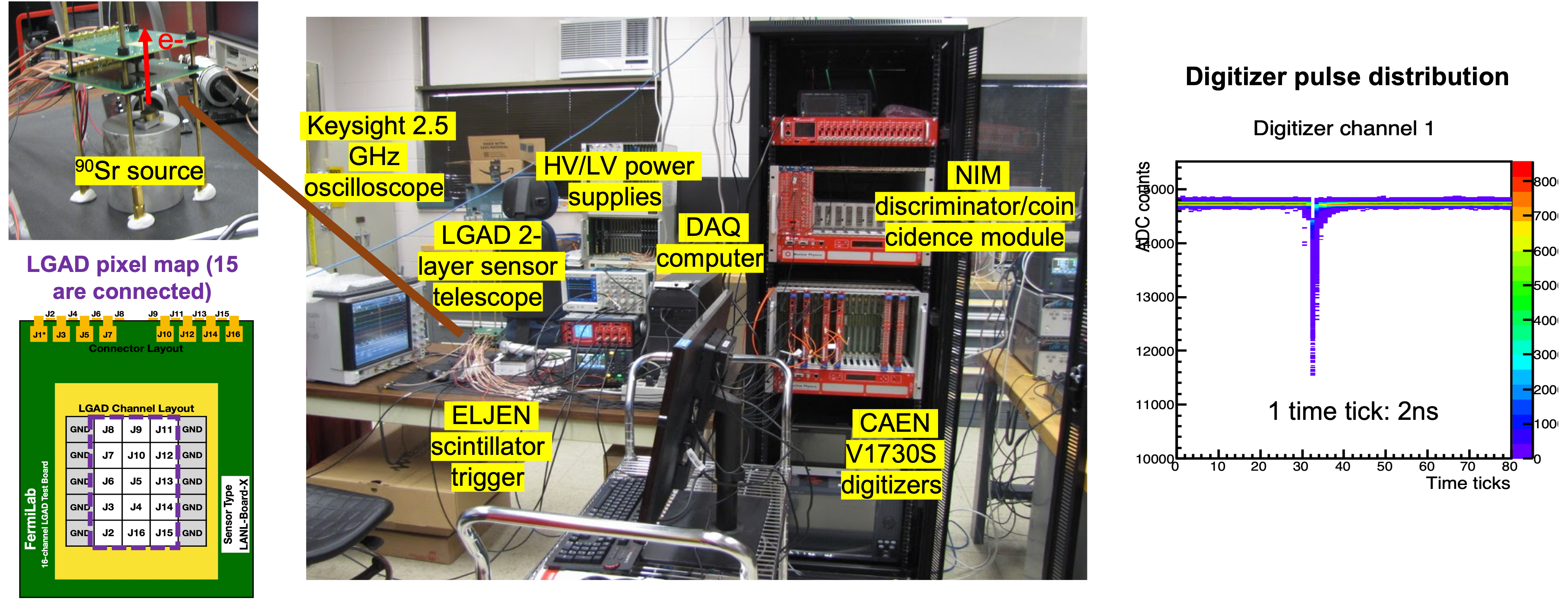}
\end{center}
\caption{Configuration of the two-layer LGAD telescope $^{90}Sr$ source test. Each LGAD sensor has 15 pixels wire bonded to the readout output (left). The analog pulse from individual pixel is shown in the right top panel. The LGAD digitized pulse processed by the CAEN 1730s digitizer is shown in the right bottom panel.}
\label{fig:lgad_bench}
\end{figure}
\begin{figure}[!htb]
\begin{center}
\includegraphics[width=0.31\textwidth]{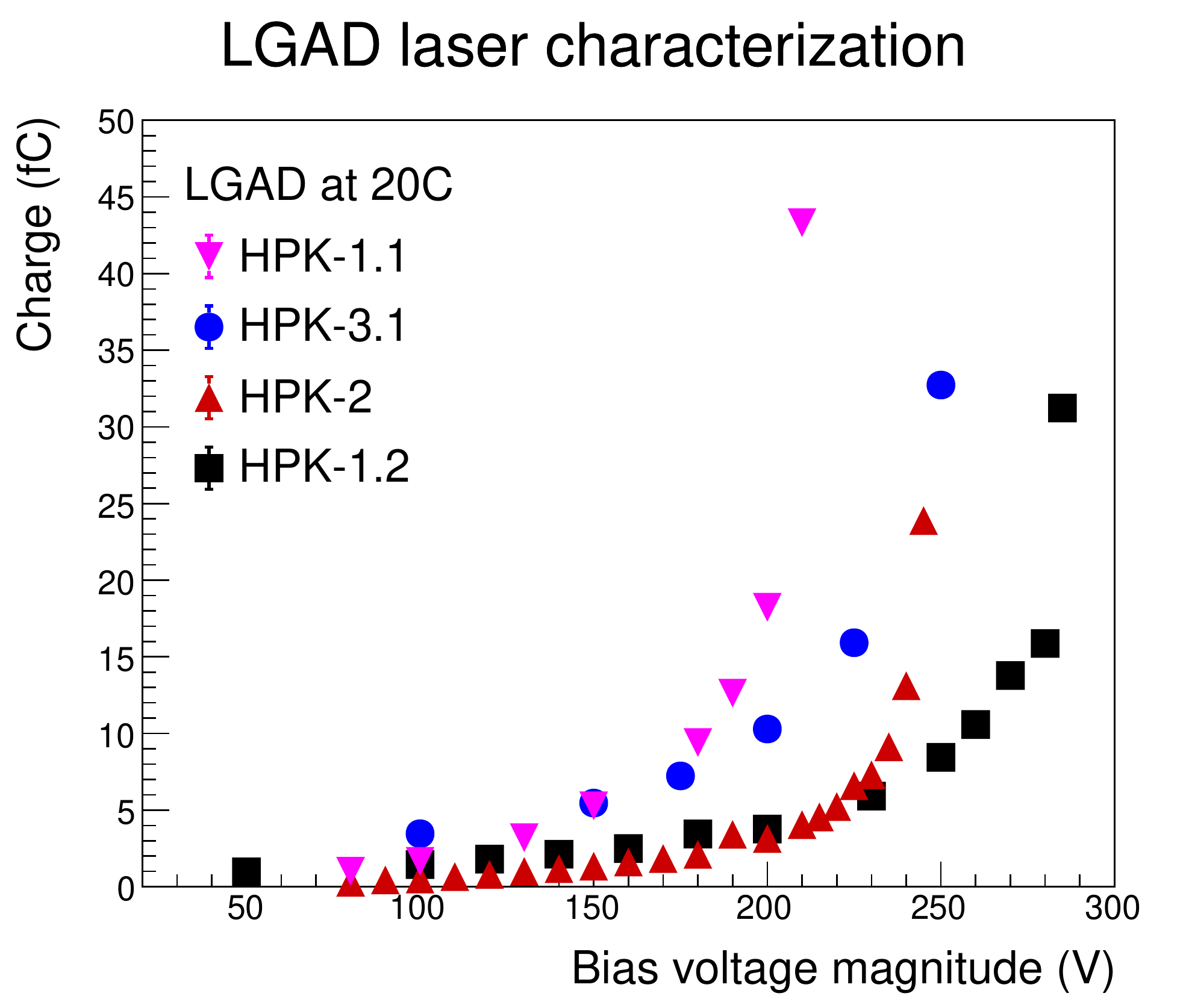}
\includegraphics[width=0.31\textwidth]{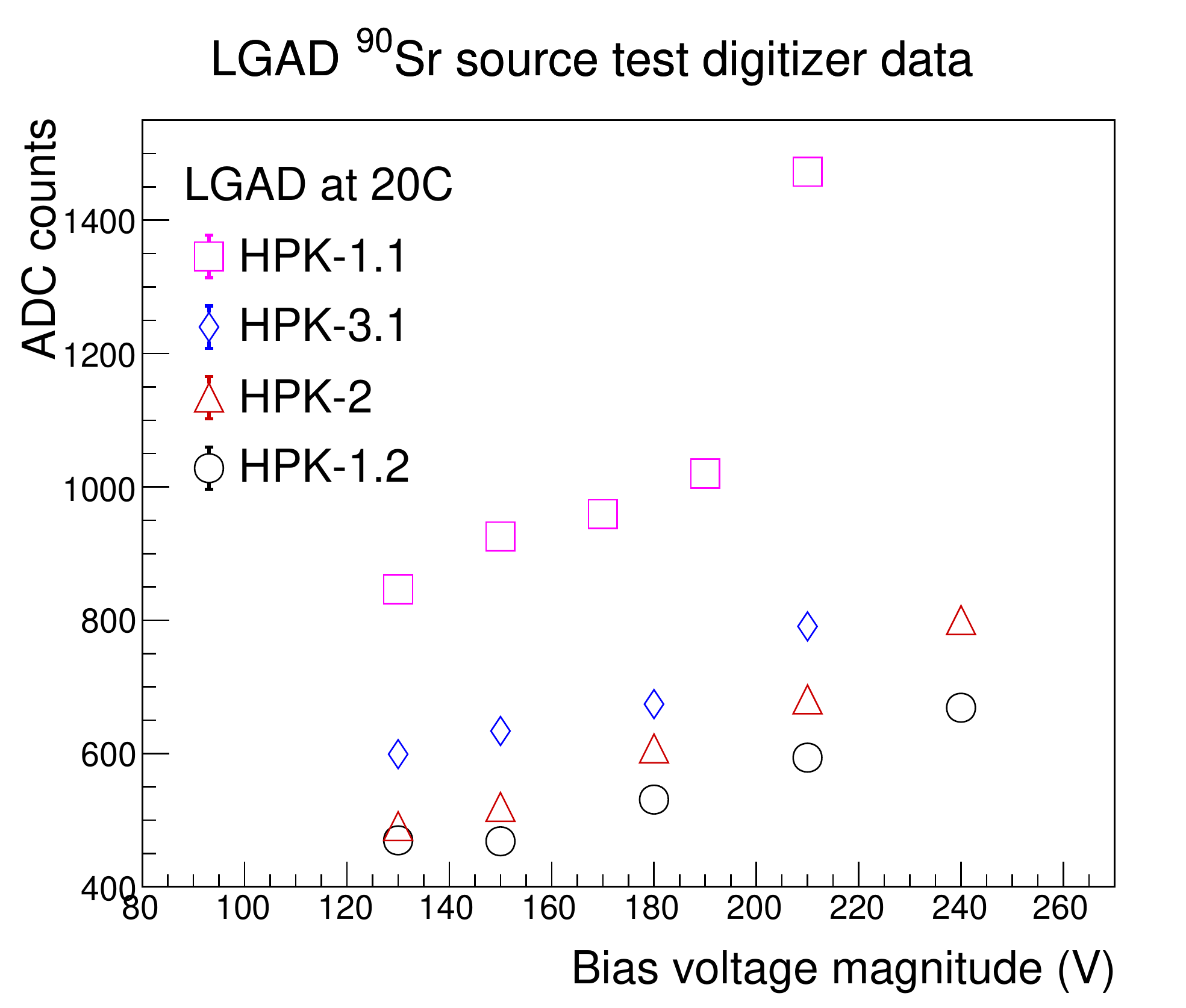}
\end{center}
\caption{The bias voltage magnitude dependent charge collections for four different types of LGAD sensors. Sensor test result with a laser scan for one set of four LGAD sensors is shown in the left and analysis results of the $^{90}Sr$ source test digitized data with an another set of four LGAD sensors are illustrated in the right.}
\label{fig:lgad_tes}
\end{figure}

\section{R$\&$D for Advanced Silicon Technology Candidates}
In addition to the MAPS and Depleted MAPS technologies, other advanced silicon sensor candidates such as Low Gain Avalanche Diode (LGAD) \cite{lgad} and AC-Coupled LGAD (AC-LGAD) \cite{ac-lgad} are under consideration for the generic EIC silicon vertex/tracking detector. These technologies can provide either better than 5 $\mu$m single hit spatial resolution or less than 30 $ps$ timing resolution. Test benches for both LGAD (AC-LGAD) and MALTA prototype sensors have been set up at Los Alamos National Laboratory. 

\begin{figure}[!htb]
\begin{center}
\includegraphics[width=0.81\textwidth]{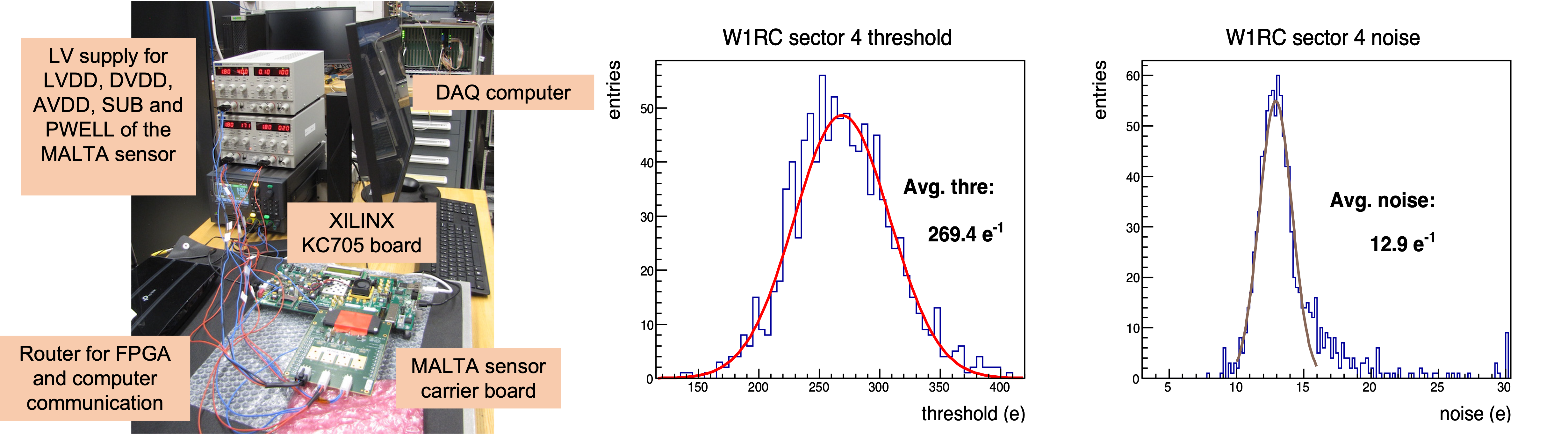}
\end{center}
\caption{Bench test configuration for a single MALTA sensor is shown in the left. Threshold and noise scan results for a partial region of a MALTA sensor are illustrated in the right.}
\label{fig:malta_tes}
\end{figure}

To characterize the LGAD (AC-LGAD) prototype sensor performance, $^{90}Sr$ source tests are used. The LGAD prototype sensor has 15 pixels wire bonded to the output of the carrier board. The LGAD pixel has a dimension of 1.3 mm by 1.3 mm, which are further reduced for AC-LGAD sensors. The readout chain based on the CAEN 1730s digitizers has been established to process the data from a single LGAD (AC-LGAD) sensor or a telescope of two LGAD (AC-LGAD) sensors with a $^{90}Sr$ source. The $^{90}Sr$ source bench test configuration for a two-layer LGAD telescope is shown in Figure~\ref{fig:lgad_bench}. The analog signal from individual pixel has a pulse width of $\sim$500 ps and pulse amplitude of $\sim$100 mV. LGAD sensors with different doping techniques have nonidentical slopes in the bias voltage dependent charge collection. The charge collection versus the bias voltage magnitude of four different types of LGAD sensors characterized with both laser scans and $^{90}Sr$ source tests are shown in Figure~\ref{fig:lgad_tes}. Similar trends have been observed for two sets of four LGAD sensors between the $^{90}Sr$ source digitized data and the analog outputs using laser beams.

The test bench to study the MALTA sensor performance has been setup and the test configuration is shown in the left panel of Figure~\ref{fig:malta_tes}. This test utilizes the XILINX KC705 FPGA evaluation board to process the data. A router, which can expand the test capability from a single sensor to a four-layer telescope, is used for the communication between the FPGA board and the DAQ computer. Each MALTA prototype sensor has a 512$\times$512 pixel matrix \cite{malta} that can be further divided into eight different regions using different pixel spacing, dopings and reset mechanisms. The pixel size of MALTA sensors is 36.4 $\mu$m by 36.4 $\mu$m. MALTA sensors can inject charges into individual pixels to evaluate their performance. Based on this feature, a series of threshold and noise scans have been carried out for different regions in a MALTA sensor. The right panel of Figure~\ref{fig:malta_tes} presents the threshold and noise scan results for 1010 pixels in the 5th region of a MALTA sensor. The average threshold over noise ratio is larger than 20, which indicates the MALTA sensor can significantly suppress the background with this threshold configuration. The noise tail due to the Random-Telegraph-Signal (RTS) has been mitigated by using a larger transistor in the new MALTA prototype (Mini-MALTA) \cite{minimalta}. Further studies such as the evaluation of hit/tracking spatial resolution and efficiency will be performed on a multi-layer telescope using a $^{90}Sr$ source and cosmic rays.

\section{Summary and Outlook}
The proposed Forward Silicon Tracker (FST) for the future EIC provides fine tracking momentum and spatial resolutions in the forward pseudorapidity region. Fast timing capability can be provided by this detector design as well. Initial characterization results of the LGAD and MALTA prototype sensors obtained from lab bench tests are consistent with their design parameters, suggesting simulations of tracking momentum resolution and DCA can be realized. Ongoing and future silicon detector R$\&$D and developments towards finer pixel pitch and faster readout speed will pave the path for the EIC detector construction and operation planned for starting in 2025.

\end{document}